# Fermi-arc diversity on surface terminations of the magnetic Weyl semimetal $Co_3Sn_2S_2$

**Authors:** Noam Morali[1]†, Rajib Batabyal[1]†, Pranab Kumar Nag[1]†, Enke Liu[2,3], Qiunan Xu[2], Yan Sun[2], Binghai Yan[1], Claudia Felser[2], Nurit Avraham[1], Haim Beidenkopf[1]*

**Affiliations:**

[1]Department of Condensed Matter Physics, Weizmann Institute of Science, Rehovot 7610001, Israel.

[2]Max Planck Institute for Chemical Physics of Solids, D-01187 Dresden, Germany.

[3]Institute of Physics, Chinese Academy of Sciences, Beijing 100190, China.

*haim.beidenkopf@weizmann.ac.il

†equal contributions

**Bulk-surface correspondence in Weyl semimetals assures the formation of topological "Fermi-arc" surface bands whose existence is guaranteed by bulk Weyl nodes. By investigating three distinct surface terminations of the ferromagnetic semimetal $Co_3Sn_2S_2$ we verify spectroscopically its classification as a time reversal symmetry broken Weyl semimetal. We show that the distinct surface potentials imposed by three different terminations modify the Fermi-arc contour and Weyl node connectivity. On the Sn surface we identify intra-Brillouin zone Weyl node connectivity of Fermi-arcs, while on Co termination the connectivity is across adjacent Brillouin zones. On the S surface Fermi-arcs overlap with non-topological bulk and surface states that ambiguate their connectivity and obscure their exact identification. By these we resolve the topologically protected electronic properties of a Weyl semimetal and its unprotected ones that can be manipulated and engineered.**

Topological semimetals lend a unique experimental opportunity to investigate both bulk and surface topological properties by probing their topological surface states *(1-6)*. Unlike in gapped topological system the dispersion of the topological surface bands is intimately correlated with the topological bulk bands dispersion *(7, 8)*. Prime examples are Weyl and Dirac semimetals whose nontrivial topological properties arise from the existence of nondegenerate band-touching points, termed Weyl nodes, in the electronic bulk band structure. Bulk Weyl nodes are formed under broken inversion or time reversal symmetry. They exhibit certain chirality and give rise to the formation of open-contour "Fermi-arc" surface bands that emanate from a certain Weyl node and terminate at another with opposite chirality within the surface two dimensional momentum space. Due to this surface-bulk correspondence, the dispersion of the Fermi-arcs reflects the Weyl cone band structure and particularly the bulk Weyl nodes energy. The Berry curvature associated with Weyl and Dirac nodes have been shown to result in unique phenomenology, most notably the chiral anomaly in magneto-transport *(9-11)* and various non-local transport effects *(12, 13)*.

The formation of Fermi-arcs is guaranteed by the existence of bulk Weyl nodes, and therefore provides a direct way to classify the bulk topology via surface probes. Nevertheless, the actual momentum space contour of the Fermi-arc bands and their connectivity among the Weyl nodes are not predetermined by the bulk distribution of Weyl nodes. Their exact shape, dispersion, and connectivity have been thought to depend on the details of the local surface potential *(14,15)*. These Fermi-arc's properties would have direct implications on various magneto-transport experiments such as quantum oscillations that uniquely combine bulk and surface conduction *(16, 17)*. Nevertheless, the susceptibility of the Fermi-arcs to surface potentials have not been studied experimentally.

Here we study the compound $Co_3Sn_2S_2$ that was recently classified as a candidate time reversal symmetry broken Weyl semimetal *(18,19)*. In contrast to inversion symmetry broken Weyl semimetals, predicted to exist in a wide variety of compounds *(6,20)* and some of which realized experimentally *(21-24)*, there are only a few candidates of magnetically ordered materials for the realization of time reversal symmetry broken Weyl semimetals. These include GdPtBi *(25)*, $Y_2Ir_2O_7$ *(1)*, $HgCr_2Se_4$ *(26)*, and certain $Co_2$-based Heusler compounds *(27)*. In few, including recently in $Co_3Sn_2S_2$ *(18,28)*, a large anomalous Hall conductivity was reported *(29)*, as an indication for the existence of a magnetic Weyl phase, though none has been spectroscopically verified.

We use scanning tunneling spectroscopy to visualize the Fermi-arcs and to investigate their structure and connectivity in $Co_3Sn_2S_2$. Its magnetic properties arise from the kagome-lattice cobalt planes (Fig. 1A), whose magnetic moments order ferromagnetically out of plane below 175 K *(30,18)*. The Co planes are interleaved with buffer planes of triangularly ordered Sn and S. *ab-initio* calculation *(18,19)* finds six Weyl nodes that form 105 meV above the chemical potential in the bulk Brillouin zone, sketched in Fig. 1B. Their projection on the (001) surface that we study here identifies three Fermi-arc bands that connect the six surface projected Weyl nodes.

The layered crystal structure of $Co_3Sn_2S_2$ enabled us to study spectroscopically all three terminations of the (001) surface. Single crystals of $Co_3Sn_2S_2$ were cold cleaved at 80 K under ultrahigh vacuum conditions, and measured at 4.2 K in a commercial STM (UNISOKU). The energetically favorable cleave plane is between the Sn and the S monolayers (Fig. 1A, see SM for further information). Indeed, most of the cleaved surface exhibits a triangular atomic structure, indicative of the Sn or S terminations, shown in Fig. 1C and D, respectively. Rarely, however, we detect the less probable Co termination, hallmarked by the characteristic Kagome crystal structure as shown in Fig. 1E. Each termination imposes a distinct surface potential that results in a distinct surface band structure. This diversity is captured by the characteristic dI/dV spectra we find on the Sn, S and Co terminated surfaces, shown in Fig. 1F, G and H, respectively.

Interestingly, our *ab-initio* calculations extend previous studies *(19)* and predict that the three different terminations of the (001) surface in $Co_3Sn_2S_2$ not only exhibit distinct Fermi-arc contours, but also distinct connectivity of the (001) surface projected Weyl nodes. While on the Sn termination the Fermi-arcs connect Weyl nodes within the same Brillouin zone (Fig. 1I), on the S termination the connectivity is ambiguated by hybridization with non-topological surface projected bulk bands (Fig. 1J), and on the Co termination the connectivity is across adjacent Brillouin zones (Fig. 1K). Studying this intricate bulk-boundaries correspondence uniquely enable us to firmly confirm the formation of a magnetic Weyl Semimetal phase in $Co_3Sn_2S_2$, and furthermore to show the susceptibility of Fermi-arcs-states in Weyl-semimetals to inherently varying surface potential. This opens the door to the manipulation of Fermi-arcs by means of engineered surface perturbations.

To visualize the Fermi-arcs and explore their energy evolution on the various surface terminations we carried out measurements of the quasi particle interference (QPI) patterns that elastically scattered electrons embed in the local density of states (LDOS) measured in differential conductance (dI/dV) maps *(31,32)*. Topographic image of the Sn terminated surface decorated by moderate concentration of adatoms is shown in Fig. 2A. The QPI patterns on the Sn surface appear as weak ripple modulations in the corresponding dI/dV map (Fig. 2B). We trace these QPI to originate from subsurface impurities (see Fig. 1C). Fourier decomposition of the QPI patterns separates surface scattering processes according to their transferred momentum, $q$, between incoming and scattered electronic wave-functions.

The QPI patterns on the Sn surface assume particularly sharp polygon shapes. Representative QPI maps taken at 7.5 meV (Fig. 2C) and at -10 meV (Fig. 2D) capture the main QPI patterns with scattering processes that involve the Fermi-arc bands. The QPI patterns at 7.5 meV (Fig. 2C) consist of a hexagon shape QPI around $q=0$ and replications of it centered on Bragg peaks. To associate these QPI patterns with particular scattering processes, we consider the calculated surface density of states DOS(k) of the Sn termination, plotted in Fig. 2E along with a simplified schematic diagram of it (left and right panels, respectively). The observed hexagonal QPI pattern originates from scattering processes (pink arrow in Fig. 2E) between a hexagonal electron pocket around $\Gamma$ and the adjacent edges of triangular electron pockets at the K and K' corners of the Brillouin zone. The edges of the triangular pockets around K are open-contour Fermi-arc bands, each of which connects a pair of Weyl cones within the Brillouin zone (note, that away from the Weyl node energy Fermi-arcs merge with the rims of the dispersing Weyl cone). The hexagonal QPI patterns therefore originate both from scattering among surface states as well as from scattering between surface to Fermi-arcs states.

With decreasing energy the hexagonal QPI patterns increase in size and towards -10 meV an additional QPI pattern appears in the form of straight lines along $\Gamma$-M that connect the adjacent corners of the QPI hexagons (Fig. 2D). The resulting QPI pattern tiles periodically q-space and accordingly signifies a pocket structure of higher symmetry. Indeed we find that the additional connecting lines in the QPI pattern originate from quasi-nesting conditions, marked by the vertical pink arrow in Fig. 2F, brought about by approximate co-linearity of edges of adjacent triangular pockets at K and K'. These scattering processes necessarily involve Fermi-arc bands. The shape, size and orientation of the connecting line QPI patterns indicate that the Fermi-arcs involved in these scattering process connect between pairs of adjacent Weyl nodes within the Brillouin zone.

The full energy evolution of the QPI patterns we find on the Sn surface is given by the momentum cut in Fig. 2G. The most prominent dispersing line (left pointing pink arrow), which curves towards $q=0$ with increasing energy, represents the hexagonal QPI patterns whose size shrinks with energy. As shown in Fig. 2E, the size of the hexagonal QPI is determined by the scattering vector between the edges of the hexagonal and triangular pockets. These pockets expand with energy (shown Fig. 2H), hence the inter pockets scattering vector and its associated hexagonal QPI decrease with energy. This evolution continues, until at around -10 meV (gray arrows in Fig. 2F and H), the K and K' pockets become comparable in size. This accidental symmetry supports the quasi-nesting conditions and the onset of the connecting lines in QPI, signified by the dark shaded area along the Γ-M direction in Fig. 2G (right pointing pink arrow) that connects the dispersing mode around $q=0$ with its replication around the Bragg peak at $q=13.5$ nm$^{-1}$.

Close inspection further reveals that, around zero bias, an inner fainter scattering mode appears along the Γ-K direction. It is hardly visible since it overlaps with the central $q=0$ broad peak that originates from inevitable long-wavelength inhomogeneities in the dI/dV map. To recover it we use the fact that these inhomogeneities are isotropic (see central disc shapes in Fig. 2C and D) thus rendering the broad $q=0$ peak symmetric along Γ-K and Γ-M. In the inset to Fig. 2G we subtract the Γ-M QPI cut from the Γ-K one. This subtraction completely eliminates the $q=0$ peak, which allows us now to resolve two dispersing scattering modes. We naturally attribute this doubled mode to scattering processes from the edge of the hexagonal pocket at Γ to the near edge of the triangular pockets at both K and K'. The distinct extent in momentum space of the triangular pockets is a direct manifestation of broken time reversal symmetry and the appearance of the second inner QPI mode is a direct measure of it. These signatures of the time reversal symmetry broken surface band structure and the Fermi-arcs it hosts on the Sn termination provide striking experimental confirmation for the existence of a magnetic Weyl semimetal phase in $Co_3Sn_2S_2$.

We now show that the Co terminated surface exhibits distinct configuration of Fermi-arcs with different connectivity than the one observed on the Sn terminated surface. A topographic image of a Co terminated surface along with its corresponding dI/dV map are shown in Fig. 3A and B, respectively. The corresponding QPI linecut taken along the high symmetry line K-Γ-M (Fig.

3C) presents a rich QPI pattern with replications that span up to the second order Bragg peaks. To recognize the main scattering processes, in particular the ones involving Fermi-arcs, we first identify the three main regimes in the energy evolution of the band structure. At low energies (Fig. 3D) we find again six closed electron pockets around the Brillion zone corners at K and K'. Upon increasing the energy, all pockets widen until the three pockets around the K points, disconnect and flatten, revealing the Fermi-arc connectivity (Fig. 1E). In contrast to the Weyl node connectivity observed on the Sn termination, where the Fermi-arcs connect a pair of Weyl nodes within the same Brillion zone, here the Fermi-arcs connect a pair of Weyl nodes from adjacent Brillion zones. These changes in Fermi-arcs connectivity are brought about by hybridization with a particular set of surface bands that coexist on the Sn surface. Upon further increase in energy the surface pockets around K' split by hybridizing with the Fermi-arc bands around K, forming C-shaped bands that are partially composed of Fermi-arc states (Fig. 1F).

These three regimes of pocket structures correspond to three regimes we identify in the evolution of the QPI patterns in Fig. 3C: At high energies (about 50-80 meV) we find a single non dispersing mode (Fig. 3G and H); At intermediate energies (about 20-50 meV) we find a single slightly dispersing mode; at low energies (-10-20 meV) we find several dispersing QPI patterns (Fig. 3I and J). Identifying this energy evolution together with theoretical calculations of the joint density of states (JDOS) allowed us to identify the scattering processes that give rise to the QPI patterns that are marked in dotted lines and arrows of corresponding color in Fig. 3C and D-F, respectively (see supplementary for further information). Among the scattering processes we have identified we concentrate here on those which involve Fermi-arcs states (pink arrows), and bands which evolve with energy into Fermi-arcs. The high energy scattering between the C-shaped bands generates the slightly dispersing high intensity QPI peaks in Fig. 3G, as well as in the corresponding JDOS (circled in both). The low energy scattering within the K pocket (pink arrow in Fig. 3D) that splits to Fermi-arc bands at slightly higher energies is identified with the diamond-like QPI pattern replicated beyond the Bragg peak in Fig. 3J (pink arrow). The QPI patterns we detect on the Co terminated surface differs dramatically from those we observed on the Sn termination. This allowed us to deduce a distinct surface band structure with distinct shape of Fermi-arcs and Weyl nodes connectivity.

We end with discussing the S terminated surface band structure which presents a third distinct case. In contrast to gapped systems where the topological surface states are protected by the bulk gap, the protection of the surface Fermi-arcs is more subtle. To be protected from scattering into the bulk, the contour of the Fermi-arcs should lie within the Weyl node gap. The local surface potential, however, is able to push the Fermi-arcs contour out of the Weyl gap and hence to lift their protection from scattering into the bulk. This is the case for the S terminated surface for which the Fermi-arcs overlap with surface projected bulk bands almost at all energies and momenta away from the Weyl node energy *(19)*. Still, the S terminated surface, whose topographic landscape is dominated by vacancies (Fig. 4A) exhibits clear signatures of electronic interference in the dI/dV mappings (Fig. 4B).

The energy dispersion of the QPI pattern on the S terminated surface along the K-Γ-M symmetry line is shown in Fig. 4C. We find several dispersing patterns (shown in details in the supplementary), however here we focus our attention to the QPI features that appear along the Γ-M direction at around 100 meV (within the dotted rectangle). In the Fourier transformed dI/dV map taken at 100 meV (Fig. 4D) these QPI features correspond to the Γ-M broad peaks, located just next to the Bragg peaks, and to the centered flower-like pattern. In the DOS(k) at the corresponding energy shown in Fig. 4E and H we identify a single dominant surface band that decorates the rims of the bulk band and terminates at the vicinity of the Weyl nodes. Detailed calculation shows that it indeed hybridizes with the Fermi-arc bands as they emanate from the Weyl nodes *(19)*. The calculated JDOS (Fig. 4F) is remarkably similar to the measured QPI, showing both the broad peaks and the flower-like patterns. This identifies the broad QPI peaks with inter-band scattering processes (bounded by $Q_a$ and $Q_A$) and the flower-like pattern with intra-band scattering processes (bounded by $Q_b$) of that surface band.

Following the energy evolution of the QPI pattern, highlighted in Fig. 4G, we find that it shifts towards smaller momentum transfers with increasing energy and seems to terminate below 200 meV. This dispersion is well captured by the DOS(k) calculation in Fig. 4H which shows that the surface band remains bound to the rims of the bulk band until it hybridizes with it at higher energies. Finally, cutting the dispersion DOS(k) through two adjacent Weyl nodes as shown in Fig. 4I reveals that the projected bands the surface state follows are the bulk Weyl cones. The dispersion of the QPI pattern in Fig. 4G accordingly embodies the dispersion of the Weyl cones.

In particular, its vanishing slightly below 200 meV corresponds to the merging of the two adjacent bulk Weyl cones and the termination of the Weyl gap. While on the S termination the Fermi-arcs contours were pushed out of this bulk gap and therefore could not be detected, on the Sn and Co terminations they lie within this gap which protects their hybridization and allows their detection.

In summary, we have detected three distinct QPI patterns on three surface terminations of $Co_3Sn_2S_2$ which confirm its topological classification as a magnetic Weyl semimetal. While all terminations share a common topological bulk band structure, each termination presents a different surface Fermi-arc dispersion. On the Sn and Co surface terminations we show distinct Fermi-arc connectivity among the surface projected Weyl nodes, while on the S termination the Fermi-arc is fully hybridized with metallic surface states such that its contour and connectivity cannot be inferred. Yet, the simple surface band structure of the S termination allowed us to extract the extent of the Weyl gap induced by the ferromagnetic order in this compound. Our results characterize the time reversal symmetry broken Weyl phase of the semimetal $Co_3Sn_2S_2$ and demonstrate the unprotected aspects of topological semimetals that allows to manipulate and engineer them by surface perturbations.

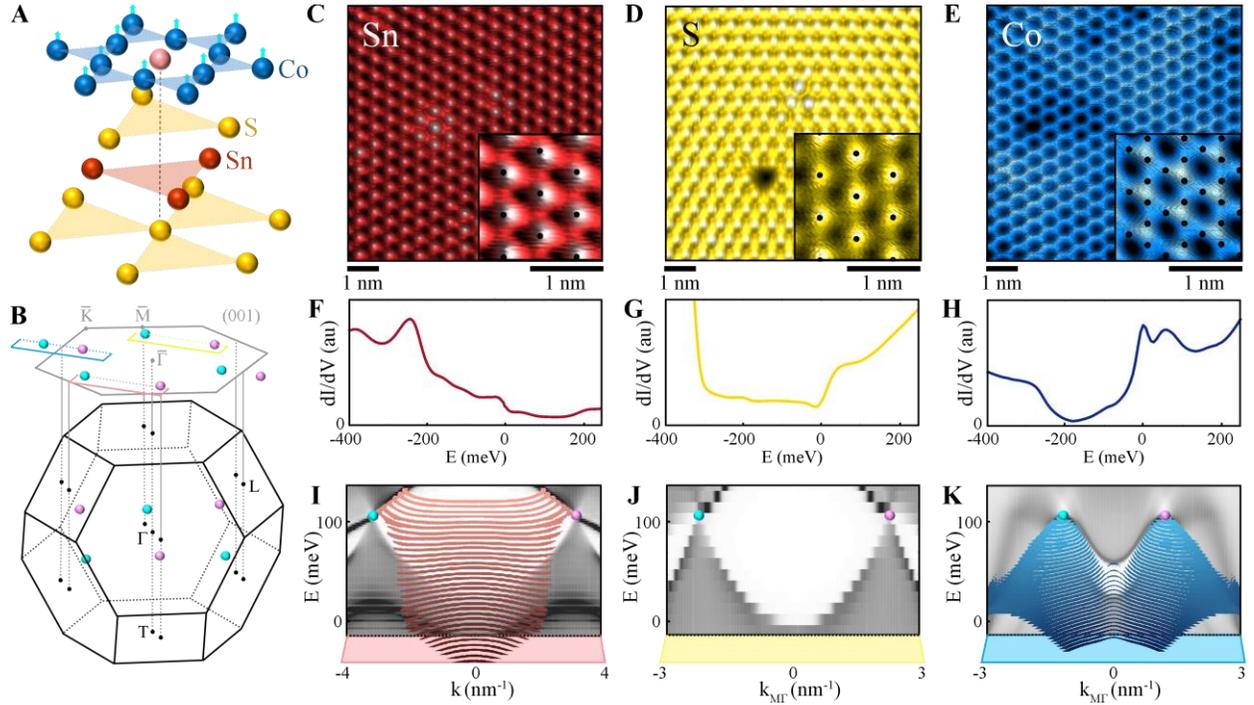

**Fig. 1.** Diverse surface band structure on Sn, S, Co terminations of $Co_3Sn_2S_2$ (A) The layered crystal structure with Co moments (arrows) ordered ferromagnetically. (B) Bulk Brillouin zone hosting three pairs of Weyl nodes and their (001) surface projection. (C) Atomically resolved Sn, S and Co surfaces, showing triangular, triangular and kagome crystal structure, respectively. Insets show atomic lattice sites indicated by black dots. (F-H) Typical dI/dV spectra on the different terminations. (I-K) *ab-initio* calculation of the band structure of $Co_3Sn_2S_2$ projected to Sn, S and Co surface terminations. A cut through a pair of Weyl nodes (along dotted line in B) is given in grayscale, while the Fermi-arc dispersions are marked in color. On Sn the Fermi arcs connect Weyl nodes within the surface Brillouin zone, on S the connectivity is obscured by metallic surface bands and on Co they connect Weyl nodes across adjacent Brillouin zones.

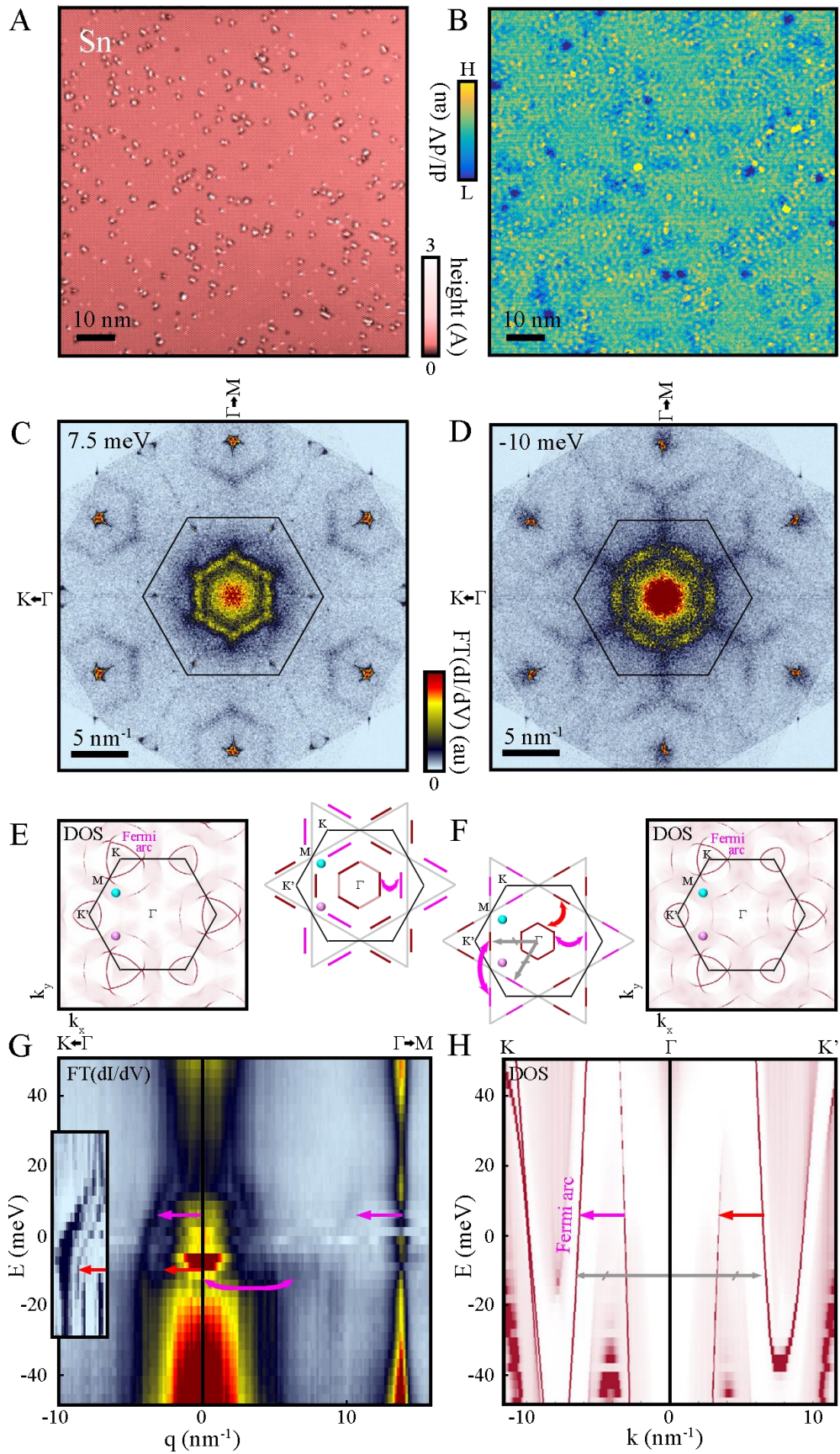

**Fig. 2.** Intra Brillouin zone Fermi arc connectivity and time reversal breaking on the Sn termination. (A) Topographic image of Sn terminated (001) surface ($V_{bias}$=95 meV, $I_{set}$=175 pA) featuring adatoms and sub-surface impurities. (B) dI/dV measurement taken on the region shown in A at $V_{bias}$=7.5 meV ($V_{AC}$=2.1 meV RMS, f=733 Hz) showing clear interference patterns. (C-D) Fourier transform of two dI/dV maps taken at different energies showing sharp QPI patterns. (E-F) DFT calculated DOS(k) and a simplified schematic diagram highlighting the important bands and scattering processes among them involving Fermi arc or surface bands (pink or red arrows, respectively). (G) Energy-Momentum cut of the QPI along K-Γ-M directions. Identified scattering processes are marked with arrows. The inset shows background eliminated two parallel modes obtained by subtracting the Γ-M cut from the Γ-K cut. (H) Energy-momentum cut of the calculated DOS along K-G-K' capturing the energy evolution of the various scattering processes marked by arrows.

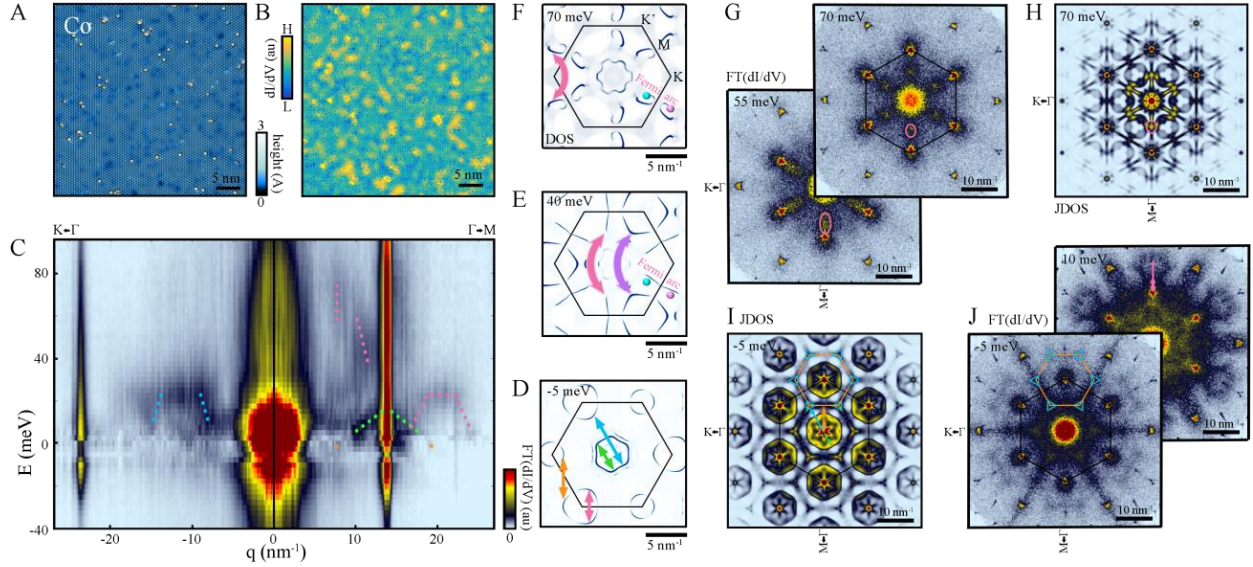

**Fig. 3.** Inter Brillouin zone Fermi arc connectivity on the Co surface. (A) Topographic image of Co terminated (001) surface ($V_{bias}$=-95 meV, $I_{set}$=250 pA) featuring adatoms and sub-surface impurities. (B) dI/dV measurement taken on the region shown in A at $V_{bias}$=-5 meV ($V_{AC}$=2.1 meV RMS, f=733 Hz). (C) Energy-momentum cut of the QPI along K-Γ-M direction. Dispersing scattering peaks are marked by dotted lines. Those that involve Fermi-arc bands appear in pink. (D-F) Representative calculated DOS(k) at different energies, signifying three major regimes in the energy evolution of the main electron pockets in the band structure. Identified scattering processes among them are marked in arrows with corresponding colors to the dotted lines in C. (G) Fourier transform of two dI/dV maps taken at relatively high energies, showing dispersive QPI broad peaks along Γ-M. (H) Calculated JDOS at the respective energy. (I) Calculated JDOS at low energy. (J) Fourier transform of two dI/dV maps taken at relatively low energies showing rich dispersing QPI patterns.

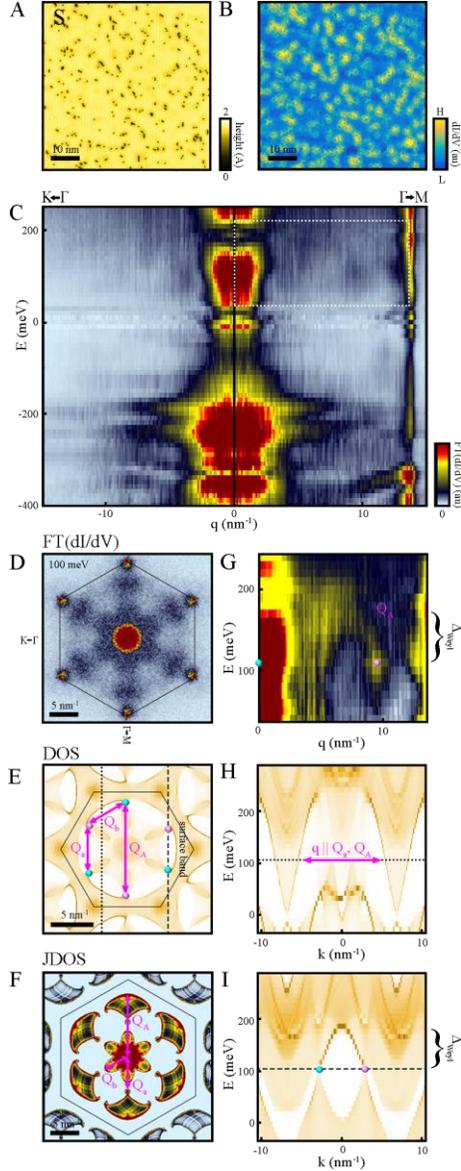

**Fig. 4.** Weyl cone dispersion derived from QPI on the S termination (A) Topographic image of the S terminated (001) surface ($V_{bias}$=50 meV, $I_{set}$=250 pA) featuring S vacancies. (B) dI/dV measurement taken on the region shown in A ($V_{bias}$=100 meV, $V_{AC}$=5.3 meV RMS, f=733 Hz), showing electronic interference patterns correlated with the location of the vacancies. (C) Energy-momentum QPI cut along the K-Γ-M direction showing several dispersing QPI branches. (D) Fourier transformed dI/dV map at the vicinity of the Weyl node energy showing QPI pattern. (E) *ab-initio* calculated DOS(k) at the corresponding energy showing a single sharp surface band and the extremal scattering wave-vectors among it. (F) Calculated JDOS at the corresponding energy agrees nicely with the measured QPI pattern in D. (G) Highlighted QPI cut (from the dotted rectangle in C) showing inward dispersing peak. (H) Energy-momentum cut of the DOS(k) along the dotted line in E showing the evolution of the scattering process. (I) Energy-momentum cut of the DOS(k) along the dashed line in E showing the extent of the Weyl gap as projected to the S surface termination.